# A problem with the Schwinger term in Dirac field theory

by

Dan Solomon

Rauland-Borg Corporation
3450 W. Oakton Street
Skokie, IL 60076
USA

Phone: 1-847-324-8337
Email: dan.Solomon@rauland.com

**Abstract**

In order for Dirac theory to be gauge invariant it can be shown that the Schwinger term must be zero. However, it can also be shown that for the vacuum state to be the lowest energy state the Schwinger term must be nonzero. Therefore there is an inconsistency in Dirac theory involving the evaluation Schwinger term. This inconsistency is discussed, along with a possible way to resolve it.

## I. Introduction

In this article we will address an inconsistency involving the Schwinger term in Dirac field theory. The Schwinger term, ST, is defined by the commutator,

$$\text{ST}(\vec{y},\vec{x}) = \left[\hat{\rho}(\vec{y}), \hat{\vec{J}}(\vec{x})\right] \tag{1}$$

where $\hat{\rho}(\vec{y})$ is the charge operator and $\hat{\vec{J}}(\vec{x})$ is the current operator. As will be shown below there is a problem with evaluating the Schwinger term. When some of the elements of Dirac theory are used it can be shown that this term must be zero. However when other elements are used then this quantity can be shown to be nonzero. Therefore there is an inconsistency in Dirac theory when it comes to evaluating this term.

Dirac field theory includes the following four elements. The first element is an evolution equation (the Schrödinger equation) which governs how the system changes in time from an initial state. In the Schrödinger representation of Dirac field theory the time evolution of the state vector $|\Omega(t)\rangle$ and its dual $\langle\Omega(t)|$ are given by,

$$\frac{\partial|\Omega(t)\rangle}{\partial t} = -i\hat{H}(t)|\Omega(t)\rangle, \quad \frac{\partial\langle\Omega(t)|}{\partial t} = i\langle\Omega(t)|\hat{H}(t) \tag{2}$$

where $\hat{H}(t)$ is the Hamiltonian operator which is given by,

$$\hat{H}(t) = \hat{H}_0 - \int \hat{\vec{J}}(\vec{x}) \cdot \vec{A}(\vec{x},t) d\vec{x} + \int \hat{\rho}(\vec{x}) A_0(\vec{x},t) d\vec{x} \tag{3}$$

In the above expression the quantities $\left(A_0(\vec{x},t), \vec{A}(\vec{x},t)\right)$ are the electric potential. In this discussion the electric potential are assumed to be unquantized, real valued functions. The quantity $\hat{H}_0$ is the free field Hamiltonian, that is, the Hamiltonian when the electric potential is zero. Note that $\hat{H}_0$, along with the current and charge operators, are time independent which is consistent with the Schrödinger picture approach. Throughout this discussion it is assumed that $|\Omega(t)\rangle$ is normalized, i.e., $\langle\Omega(t)|\Omega(t)\rangle = 1$. Note that Eq. (2) ensures that the normalization of $|\Omega(t)\rangle$ is constant in time.

A second element of Dirac theory is the principal of gauge invariance. An important requirement of a physical theory is that it be gauge invariant. The electromagnetic field is given in terms of the electric potential by,

$$\vec{E} = -\left(\frac{\partial \vec{A}}{\partial t} + \vec{\nabla} A_0\right); \quad \vec{B} = \vec{\nabla} \times \vec{A} \tag{4}$$

A change in the gauge is a change in the electric potential which produces no change in the electromagnetic field. Such a change is given by,

$$\vec{A} \to \vec{A}' = \vec{A} - \vec{\nabla}\chi, \quad A_0 \to A_0' = A_0 + \frac{\partial \chi}{\partial t} \tag{5}$$

where $\chi(\vec{x}, t)$ is an arbitrary real valued function.

Now when a change in the gauge is introduced into (2) this will produce a change in the state vector $|\Omega\rangle$. However, for Dirac field theory to be gauge invariant a change in the gauge must produce no change in the physical observables. These include the current and charge expectation values which are defined by,

$$\vec{J}_e(\vec{x}, t) = \langle\Omega(t)|\hat{\vec{J}}(\vec{x})|\Omega(t)\rangle \text{ and } \rho_e(\vec{x}, t) = \langle\Omega(t)|\hat{\rho}(\vec{x})|\Omega(t)\rangle \tag{6}$$

A third element that we expect Dirac theory to obey is that of local conservation of electric charge, that is, the continuity equation holds,

$$\frac{\partial \rho_e(\vec{x},t)}{\partial t} = -\vec{\nabla} \cdot \vec{J}_e(\vec{x},t) \tag{7}$$

The fourth element of Dirac theory to be considered in this discussion is that there exists a minimum value to the free field energy. The free field energy, $\xi_0(|\Omega\rangle)$, of a state vector $|\Omega\rangle$, is the energy of the quantum state in the absence of interactions, i.e., the electric potential is zero. $\xi_0(|\Omega\rangle)$ is defined by,

$$\xi_0(|\Omega\rangle) = \langle\Omega|\hat{H}_0|\Omega\rangle \tag{8}$$

Let $|n\rangle$ be the entire set of eigenstates of $\hat{H}_0$ with eigenvalues $\varepsilon_n$. The $|n\rangle$ form a set of basis states (see Chapt. 3 of Itzykson and Zuber[1] ) and satisfy the equations

$$\hat{H}_o|n\rangle = \varepsilon_n|n\rangle; \quad \langle n|\hat{H}_o = \langle n|\varepsilon_n \tag{9}$$

and

$$\langle n|m\rangle = \delta_{nm} \tag{10}$$

Also we can define the relationship

$$\sum_n |n\rangle\langle n| = 1 \tag{11}$$

(see Chapter VII of Messiah [2]).

The vacuum state $|0\rangle$ is generally assumed to be the eigenvector of $\hat{H}_0$ with the smallest eigenvalue $\varepsilon_o = 0$. For all other eigenvalues,

$$\varepsilon_n > \varepsilon_o = 0 \text{ for } |n\rangle \neq |0\rangle \tag{12}$$

From the above we obtain,

$$\xi_0(|\Omega\rangle) > \xi_0(|0\rangle) = 0 \text{ for all } |\Omega\rangle \neq |0\rangle \tag{13}$$

Therefore the vacuum state is the quantum state with the minimum value of the free field energy.

## II. The Schwinger term

The question that we will address in this section is whether or not these four elements of Dirac theory are mathematically consistent. From the equations of Section I we can derive a number of additional relationships, including those involving the Schwinger term. First consider the time derivative of the current expectation value. From (6) and (2) we obtain,

$$\frac{\partial \vec{J}_e(\vec{x},t)}{\partial t} = i\left\langle \Omega(t) \left| \left[ \hat{H}(t), \hat{\vec{J}}(\vec{x}) \right] \right| \Omega(t) \right\rangle \tag{14}$$

Use (3) in the above to yield,

$$\frac{\partial \vec{J}_e(\vec{x},t)}{\partial t} = i\left\langle \Omega(t) \left| \begin{pmatrix} \left[\hat{H}_0, \hat{\vec{J}}(\vec{x})\right] - \int \left[\hat{\vec{J}}(\vec{y}) \cdot \vec{A}(\vec{y},t), \hat{\vec{J}}(\vec{x})\right] d\vec{y} \\ + \int \left[\hat{\rho}(\vec{y}), \hat{\vec{J}}(\vec{x})\right] A_0(\vec{y},t) d\vec{y} \end{pmatrix} \right| \Omega(t) \right\rangle \tag{15}$$

Next perform the gauge transformation (5) to obtain,

$$\frac{\partial \vec{J}_e(\vec{x},t)}{\partial t} = i\left\langle \Omega(t) \left| \begin{pmatrix} \left[\hat{H}_0, \hat{\vec{J}}(\vec{x})\right] - \int \left[\hat{\vec{J}}(\vec{y}) \cdot \left(\vec{A}(\vec{y},t) - \vec{\nabla}\chi(\vec{y},t)\right), \hat{\vec{J}}(\vec{x})\right] d\vec{y} \\ + \int \left[\hat{\rho}(\vec{y}), \hat{\vec{J}}(\vec{x})\right] \left( A_0(\vec{y},t) + \frac{\partial \chi(\vec{y},t)}{\partial t} \right) d\vec{y} \end{pmatrix} \right| \Omega(t) \right\rangle$$

(16)

The quantity $\partial \vec{J}_e / \partial t$ is a physical observable and therefore, if the theory is gauge invariant, must not depend on the quantity $\chi$ or $\partial \chi / \partial t$. Now, at a particular instant of time $\partial \chi / \partial t$ can be varied in an arbitrary manner without changing the values of any of the other quantities on the right hand side of the equals sign in the above equation. Therefore for $\partial \vec{J}_e/\partial t$ to be independent of $\partial \chi/\partial t$ we must have that,

$$\mathrm{ST}(\vec{y},\vec{x}) = \left[\hat{\rho}(\vec{y}), \hat{\vec{J}}(\vec{x})\right] = 0 \qquad (17)$$

This relationship, and others, can also be derived from the continuity equation. To show this use (6) in (7) to obtain,

$$\frac{\partial \langle \Omega(t)|\hat{\rho}(\vec{x})|\Omega(t)\rangle}{\partial t} = -\langle \Omega(t)|\vec{\nabla}\cdot\hat{\vec{J}}(\vec{x})|\Omega(t)\rangle \qquad (18)$$

Next use (2) in the above to yield,

$$i\langle \Omega(t)|\left[\hat{H},\hat{\rho}(\vec{x})\right]|\Omega(t)\rangle = -\langle \Omega(t)|\vec{\nabla}\cdot\hat{\vec{J}}(\vec{x})|\Omega(t)\rangle \qquad (19)$$

Use (3) in the above to yield,

$$i\langle \Omega(t)|\left(\begin{array}{l}\left[\hat{H}_0,\hat{\rho}(\vec{x})\right] - \int\left[\hat{\vec{J}}(\vec{y}),\hat{\rho}(\vec{x})\right]\cdot\vec{A}(\vec{y},t)d\vec{y} \\ +\int\left[\hat{\rho}(\vec{y}),\hat{\rho}(\vec{x})\right]A_0(\vec{y},t)d\vec{y}\end{array}\right)|\Omega(t)\rangle = -\langle \Omega(t)|\vec{\nabla}\cdot\hat{\vec{J}}(\vec{x})|\Omega(t)\rangle$$

(20)

For the above equation to be true for arbitrary values of the state vector $|\Omega(t)\rangle$ and the electric potential $(A_0(\vec{x},t), \vec{A}(\vec{x},t))$ equation (17) must hold along with the additional relationships,

$$i\left[\hat{H}_0,\hat{\rho}(\vec{x})\right] = -\vec{\nabla}\cdot\hat{\vec{J}}(\vec{x}) \qquad (21)$$

$$\left[\hat{\rho}(\vec{y}),\hat{\rho}(\vec{x})\right] = 0 \qquad (22)$$

Therefore we have shown, that if we take the Schrödinger equation and assume either gauge invariance or local charge conservation then the Schwinger term must be zero.

However, there is a problem with this result. Schwinger [3] has shown that it violates element 4 of the previous section, that is, the assumption that the vacuum state is the state with the lowest free field energy state. This will be demonstrated below.

First take the divergence of the Schwinger term $\left[\hat{\rho}(\vec{y}),\hat{J}(\vec{x})\right]$ and use (21) to obtain,

$$\vec{\nabla}_{\vec{x}}\cdot\left[\hat{\rho}(\vec{y}),\hat{\vec{J}}(\vec{x})\right]=\left[\hat{\rho}(\vec{y}),\vec{\nabla}\cdot\hat{\vec{J}}(\vec{x})\right]=-i\left[\hat{\rho}(\vec{y}),\left[\hat{H}_0,\hat{\rho}(\vec{x})\right]\right] \tag{23}$$

Next expand the commutator to yield,

$$i\vec{\nabla}_{\vec{x}}\cdot\left[\hat{\rho}(\vec{y}),\hat{\vec{J}}(\vec{x})\right]=-\hat{H}_0\hat{\rho}(\vec{x})\hat{\rho}(\vec{y})+\hat{\rho}(\vec{x})\hat{H}_0\hat{\rho}(\vec{y})+\hat{\rho}(\vec{y})\hat{H}_0\hat{\rho}(\vec{x})-\hat{\rho}(\vec{y})\hat{\rho}(\vec{x})\hat{H}_0 \tag{24}$$

Sandwich the above expression between the state vector $\langle 0|$ and its dual $|0\rangle$ and use $\hat{H}_0|0\rangle=0$ and $\langle 0|\hat{H}_0=0$ to obtain,

$$i\vec{\nabla}_{\vec{x}}\cdot\langle 0|\left[\hat{\rho}(\vec{y}),\hat{\vec{J}}(\vec{x})\right]|0\rangle=\langle 0|\hat{\rho}(\vec{x})\hat{H}_0\hat{\rho}(\vec{y})|0\rangle+\langle 0|\hat{\rho}(\vec{y})\hat{H}_0\hat{\rho}(\vec{x})|0\rangle \tag{25}$$

Next set $\vec{y}=\vec{x}$ to obtain,

$$i\vec{\nabla}_{\vec{x}}\cdot\langle 0|\left[\hat{\rho}(\vec{y}),\hat{\vec{J}}(\vec{x})\right]|0\rangle\Big|_{\vec{y}=\vec{x}}=2\langle 0|\hat{\rho}(\vec{x})\hat{H}_0\hat{\rho}(\vec{x})|0\rangle \tag{26}$$

Use (11) in the above to obtain,

$$i\vec{\nabla}_{\vec{x}}\cdot\langle 0|\left[\hat{\rho}(\vec{y}),\hat{\vec{J}}(\vec{x})\right]|0\rangle\Big|_{\vec{y}=\vec{x}}=2\sum_{n,m}\langle 0|\hat{\rho}(\vec{x})|n\rangle\langle n|\hat{H}_0|m\rangle\langle m|\hat{\rho}(\vec{x})|0\rangle \tag{27}$$

Next use (9) and (10) to obtain,

$$i\vec{\nabla}_{\vec{x}}\cdot\langle 0|\left[\hat{\rho}(\vec{y}),\hat{\vec{J}}(\vec{x})\right]|0\rangle\Big|_{\vec{y}=\vec{x}}=2\sum_{n}\varepsilon_n\langle 0|\hat{\rho}(\vec{x})|n\rangle\langle n|\hat{\rho}(\vec{x})|0\rangle=2\sum_{n}\varepsilon_n\left|\langle 0|\hat{\rho}(\vec{x})|n\rangle\right|^2 \tag{28}$$

Now, in general, the quantity $\langle 0|\hat{\rho}(\vec{x})|n\rangle$ is not zero [3] and since $\varepsilon_n \geq 0$ the above expression is non-zero and positive. Therefore the Schwinger term cannot be zero. This is, of course, in direct contradiction to (17).

Therefore we see that the four elements of Dirac field theory do not produce consistent results. On one hand using the Schrödinger equation along with the principle of gauge invariance or the continuity equation we can prove that the Schwinger term is zero. On the other hand if there is a lower bound to the free field energy then the Schwinger term cannot be zero.

### III. The vacuum state

In order to evaluate the Schwinger term we must express the current and charge operators in terms of the field operators $\hat{\psi}(\vec{x})$,

$$\hat{\vec{J}}(\vec{x}) = q\hat{\psi}^\dagger(\vec{x})\vec{\alpha}\hat{\psi}(\vec{x}) \text{ and } \hat{\rho}(\vec{x}) = q\hat{\psi}^\dagger(\vec{x})\hat{\psi}(\vec{x}) \qquad (29)$$

The free field Hamiltonian operator $\hat{H}_0$ is given by,

$$\hat{H}_0 = \int \hat{\psi}^\dagger(\vec{x}) H_0 \hat{\psi}(\vec{x}) d\vec{x} - \xi_r \qquad (30)$$

where $\xi_r$ is a renormalization constant which is defined in such a way to make the free field energy of the vacuum state equal to zero.

In the above expressions 'q' is the electric charge and,

$$\hat{\psi}(\vec{x}) = \sum_n \hat{a}_n \phi_n(\vec{x}); \quad \hat{\psi}^\dagger(\vec{x},t) = \sum_n \hat{a}_n^\dagger \phi_n^\dagger(\vec{x}) \qquad (31)$$

where the $\hat{a}_n$ ($\hat{a}_n^\dagger$) are the destruction(creation) operators for a particle in the state $\phi_n(\vec{x},t)$. They satisfy the anticommutator relation

$$\hat{a}_m \hat{a}_n^\dagger + \hat{a}_n^\dagger \hat{a}_m = \delta_{mn}; \text{ all other anticommutators=0} \qquad (32)$$

The $\phi_n(\vec{x})$ are basis state solutions of the free field Dirac equation with energy eigenvalue $\lambda_n E_n$ and can be expressed by

$$H_0 \phi_n(\vec{x}) = \lambda_n E_n \phi_n(\vec{x}) \tag{33}$$

where,

$$H_0 = -i\vec{\alpha} \cdot \vec{\nabla} + \beta m \tag{34}$$

and where,

$$E_n = +\sqrt{\vec{p}_n^2 + m^2}, \quad \lambda_n = \begin{cases} +1 \text{ for a positive energy state} \\ -1 \text{ for a negative energy state} \end{cases} \tag{35}$$

where $\vec{p}_n$ is the momentum of the state n. Note that in the above equations we use $\hbar = c = 1$.

The $\phi_n(\vec{x})$ can be expressed by,

$$\phi_n(\vec{x}) = u_n e^{i\vec{p} \cdot \vec{x}} \tag{36}$$

where $u_n$ is a constant 4-spinor. The $\phi_n(\vec{x})$ form a complete orthonormal basis in Hilbert space and satisfy

$$\int \phi_n^\dagger(\vec{x},t) \phi_m(\vec{x},t) d\vec{x} = \delta_{mn} \tag{37}$$

and

$$\sum_n \left(\phi_n^\dagger(\vec{x})\right)_{(a)} \left(\phi_n(\vec{y})\right)_{(b)} = \delta_{ab} \delta^3(\vec{x}-\vec{y}) \tag{38}$$

where "a" and "b" are spinor indices (see page 107 of Heitler [4]).

Following Greiner [5] define the state vector $|0, \text{bare}\rangle$ which is the state vector that is empty of all particles, i.e.,

$$\hat{a}_n |0, \text{bare}\rangle = 0 \text{ for all n} \tag{39}$$

The vacuum state vector $|0\rangle$ is defined as the state vector in which all negative energy states are occupied by a single particle. Therefore

$$|0\rangle = \prod_{n;\lambda_n=-1} \hat{a}_n^\dagger |0,\text{bare}\rangle \qquad (40)$$

where the notation $n; \lambda_n = -1$ means that the product is taken over all negative energy states. From this expression and Eqs. (32) and (39), $|0\rangle$ can then be defined by

$$\hat{a}_n |0\rangle = 0 \text{ for } \lambda_n = 1 \; ; \; \hat{a}_n^\dagger |0\rangle = 0 \text{ for } \lambda_n = -1 \qquad (41)$$

(Note, it is possible for the negative energy states, where $\lambda_n = -1$, to replace the electron destruction and creation operators $\hat{a}_n$ and $\hat{a}_n^\dagger$ with positron creation and destruction operators $\hat{b}_n^\dagger$ and $\hat{b}_n$, respectively. However in this article it will be useful to stay with the present notation).

New state vectors can be produced by operating on $|0\rangle$ with the operators $\hat{a}_n$ (if $\lambda_n = -1$) and $\hat{a}_n^\dagger$ (if $\lambda_n = +1$). The effect of doing this is to either remove an electron from an already occupied negative energy state or to place an electron into an unoccupied positive energy state. In either case the energy of the state is increased. Based on this it is easy to show that the vacuum state $|0\rangle$ is a lower bound to the free field energy and equation (13) is true.

## IV. Evaluating the Schwinger term

Using the results of the previous section we want to determine if the Schwinger term is zero or non-zero. According to Schwinger [3] the quantity $\left[\hat{\rho}(\vec{y}), \hat{J}(\vec{x})\right]$ is a c-number therefore,

$$\text{ST}(\vec{y},\vec{x}) = \left[\hat{\rho}(\vec{y}),\hat{J}(\vec{x})\right] = \left\langle 0\left|\left[\hat{\rho}(\vec{y}),\hat{J}(\vec{x})\right]\right|0\right\rangle \quad (42)$$

Use (29) in the above to yield,

$$\text{ST}(\vec{y},\vec{x}) = q^2 \left\langle 0\left|\left[\hat{\psi}^\dagger(\vec{y})\hat{\psi}(\vec{y}),\hat{\psi}^\dagger(\vec{x})\vec{\alpha}\hat{\psi}(\vec{x})\right]\right|0\right\rangle \quad (43)$$

Use (31) in the above to obtain,

$$\text{ST}(\vec{y},\vec{x}) = q^2 \sum_{nmrs} \left\langle 0\left|\left[\hat{a}_n^\dagger \hat{a}_m, \hat{a}_r^\dagger \hat{a}_s\right]\right|0\right\rangle \left(\phi_n^\dagger(\vec{y})\phi_m(\vec{y})\right)\left(\phi_r^\dagger(\vec{x})\vec{\alpha}\phi_s(\vec{x})\right) \quad (44)$$

Use (32) to obtain,

$$\left[\hat{a}_n^\dagger \hat{a}_m, \hat{a}_r^\dagger \hat{a}_s\right] = -\delta_{ns}\hat{a}_r^\dagger \hat{a}_m + \delta_{rm}\hat{a}_n^\dagger \hat{a}_s \quad (45)$$

Next use (41) in the above to obtain,

$$\left\langle 0\left|\left[\hat{a}_n^\dagger \hat{a}_m, \hat{a}_r^\dagger \hat{a}_s\right]\right|0\right\rangle = \begin{cases} \delta_{ns}\delta_{rm} & \text{if } \lambda_m = 1 \text{ and } \lambda_s = -1 \\ -\delta_{ns}\delta_{rm} & \text{if } \lambda_m = -1 \text{ and } \lambda_s = 1 \\ 0 & \text{if } \lambda_m = -1 \text{ and } \lambda_s = -1 \\ 0 & \text{if } \lambda_m = 1 \text{ and } \lambda_s = 1 \end{cases} \quad (46)$$

Use this result in (44) to obtain,

$$\text{ST}(\vec{y},\vec{x}) = \left\{q^2 \sum_{n;\lambda_n=-1} \sum_{m;\lambda_m=+1} \left(\phi_n^\dagger(\vec{y})\phi_m(\vec{y})\right)\left(\phi_m^\dagger(\vec{x})\vec{\alpha}\phi_n(\vec{x})\right)\right\} - (\text{c.c.}) \quad (47)$$

where the expression c.c. means to take the complex conjugate of the previous term and the summations are over all negative energy states for the n indices and over all positive energy states for the m indices.

Next take the divergence of the above expression with respect to $\vec{x}$ to obtain,

$$\vec{\nabla}_{\vec{x}} \cdot \text{ST}(\vec{y},\vec{x}) = \left\{q^2 \sum_{n;\lambda_n=-1} \sum_{m;\lambda_m=+1} \left(\phi_n^\dagger(\vec{y})\phi_m(\vec{y})\right)\vec{\nabla} \cdot \left(\phi_m^\dagger(\vec{x})\vec{\alpha}\phi_n(\vec{x})\right)\right\} - (\text{c.c.}) \quad (48)$$

To evaluate this expression use the following,

$$\vec{\nabla} \cdot \left( \phi_m^\dagger(\vec{x}) \vec{\alpha} \phi_n(\vec{x}) \right) = \left( \left( \vec{\alpha} \cdot \vec{\nabla} \phi_m(\vec{x}) \right)^\dagger \phi_n(\vec{x}) \right) + \left( \phi_m^\dagger(\vec{x}) \vec{\alpha} \cdot \vec{\nabla} \phi_n(\vec{x}) \right) \tag{49}$$

Next use (34) in the above to obtain,

$$\vec{\nabla} \cdot \left( \phi_m^\dagger(\vec{x}) \vec{\alpha} \phi_n(\vec{x}) \right) = \left( \left( iH_0 \phi_m(\vec{x}) \right)^\dagger \phi_n(\vec{x}) \right) + \left( \phi_m^\dagger(\vec{x}) iH_0 \phi_n(\vec{x}) \right) \tag{50}$$

Next use (33) to obtain,

$$\vec{\nabla} \cdot \left( \phi_m^\dagger(\vec{x}) \vec{\alpha} \phi_n(\vec{x}) \right) = i\phi_m^\dagger(\vec{x})(\lambda_n E_n - \lambda_m E_m)\phi_n(\vec{x}) \tag{51}$$

Use this in (48) to yield,

$$\vec{\nabla}_{\vec{x}} \cdot ST(\vec{y},\vec{x}) = \left\{ iq^2 \sum_{n;\lambda_n=-1} \sum_{m;\lambda_m=+1} \left( \phi_n^\dagger(\vec{y})\phi_m(\vec{y}) \right)\left( \phi_m^\dagger(\vec{x})\phi_n(\vec{x}) \right)(\lambda_n E_n - \lambda_m E_m) \right\} - (c.c.)$$

$$\tag{52}$$

Use the fact that $\lambda_n = -1$ and $\lambda_m = 1$ and set $\vec{y} = \vec{x}$ to obtain,

$$\vec{\nabla}_{\vec{x}} \cdot ST(\vec{y},\vec{x}) \Big|_{\vec{y}=\vec{x}} = -2 \left\{ iq^2 \sum_{n;\lambda_n=-1} \sum_{m;\lambda_m=+1} \left( \phi_n^\dagger(\vec{x})\phi_m(\vec{x}) \right)\left( \phi_m^\dagger(\vec{x})\phi_n(\vec{x}) \right)(E_n + E_m) \right\}$$

$$\tag{53}$$

This yields,

$$\vec{\nabla}_{\vec{x}} \cdot ST(\vec{y},\vec{x}) \Big|_{\vec{y}=\vec{x}} = -2 \left\{ iq^2 \sum_{n;\lambda_n=-1} \sum_{m;\lambda_m=+1} \left| \phi_m^\dagger(\vec{x})\phi_n(\vec{x}) \right|^2 (E_n + E_m) \right\} \tag{54}$$

Each term in the sum is positive so that the above expression cannot be equal to zero. Therefore the Schwinger term $ST(\vec{y},\vec{x})$ is nonzero. This result is expected because, as explained in Section II, if the Schwinger term is zero there must be state vectors with a lower free field energy than the vacuum state, however, the vacuum state vector $|0\rangle$ obeys the relationship (13) therefore the Schwinger term must be nonzero.

## V. Gauge invariance and the Schwinger term

The problem with the result of the previous section is that, per the discussion in Section II, if the Schwinger term is nonzero then Dirac theory is not gauge invariant and the continuity equation is not valid. The problem shows up in Dirac field theory when the vacuum polarization tensor is calculated. It is well known that when the vacuum polarization tensor is calculated, using perturbation theory, the result is not gauge invariant (see Chapter 14 of [5], Sect. 22 of [4], Chapter 5 of [6], and [7]). The non-gauge invariant terms must be removed from the results of the calculation in order to obtain a physically correct result. This may involve some form of regularization, where other functions are introduced that happen to have the correct behavior so that the non-gauge invariant terms are cancelled. However there is no physical explanation for introducing these functions [7]. They are mathematical devices used to force the desired gauge invariant result.

The question is why do these non-gauge invariant terms appear in a theory that is supposed to be gauge invariant? It will be shown that this problem is directly related to the evaluation of the Schwinger term. This problem was originally addressed in [8]. The following analysis closely follows this reference.

From eq. 8.3 of [9] the first order change in the vacuum current due to an external electric field is given by,

$$\vec{J}^{(1)}_{vac}(\vec{x},t) = i\langle 0| \left[ \hat{\vec{J}}(\vec{x},t), \int d\vec{y} \int_{-\infty}^{t} dt' \left( -\hat{\vec{J}}(\vec{y},t')\cdot\vec{A}(\vec{y},t') + \hat{\rho}(\vec{y},t')A_o(\vec{y},t') \right) \right] |0\rangle \qquad (55)$$

In the above expression the operators $\hat{\vec{J}}(\vec{x},t)$ and $\hat{\rho}(\vec{x},t)$ are the current and charge operators, respectively, in the interaction representation. They are related to the Schrödinger operators, $\hat{\vec{J}}(\vec{x})$ and $\hat{\rho}(\vec{x})$, by

$$\hat{\vec{J}}(\vec{x},t) = e^{i\hat{H}_o t}\hat{\vec{J}}(\vec{x})e^{-i\hat{H}_o t} \text{ and } \hat{\rho}(\vec{x},t) = e^{i\hat{H}_o t}\hat{\rho}(\vec{x})e^{-i\hat{H}_o t} \tag{56}$$

According to Eq. 3.11 of [9] the above interaction operators satisfy the continuity equation,

$$\frac{\partial \hat{\rho}(\vec{x},t)}{\partial t} = -\vec{\nabla}\cdot\hat{\vec{J}}(\vec{x},t) \tag{57}$$

The change in the vacuum current $\delta\vec{J}^{(1)}_{vac}(\vec{x},t)$ due to a gauge transformation is obtained by using Eq. (5) in (55) to yield

$$\delta\vec{J}^{(1)}_{vac}(\vec{x},t) = i\langle 0|\left[\hat{\vec{J}}(\vec{x},t), \int d\vec{y}\int_{-\infty}^{t} dt'\left(\hat{\vec{J}}(\vec{y},t')\cdot\vec{\nabla}\chi(\vec{y},t')+\hat{\rho}(\vec{y},t')\frac{\partial\chi(\vec{y},t')}{\partial t'}\right)\right]|0\rangle \tag{58}$$

If Dirac field theory is gauge invariant then a gauge transformation should produce no change in any observable quantity. Therefore $\delta\vec{J}^{(1)}_{vac}(\vec{x},t)$ should be zero. To verify this we will solve the above equation as follows. First consider the following relationship,

$$\int_{-\infty}^{t} dt'\hat{\rho}(\vec{y},t')\frac{\partial\chi(\vec{y},t')}{\partial t'} = \left.\hat{\rho}(\vec{y},t')\chi(\vec{y},t')\right|_{-\infty}^{t} - \int_{-\infty}^{t} dt'\chi(\vec{y},t')\frac{\partial\hat{\rho}(\vec{y},t')}{\partial t'} \tag{59}$$

Assume that $\chi(\vec{y},t) = 0$ at $t\to-\infty$. Use this and Eq. (57) in the above expression to obtain

$$\int_{-\infty}^{t} dt'\hat{\rho}(\vec{y},t')\frac{\partial\chi(\vec{y},t')}{\partial t'} = \hat{\rho}(\vec{y},t)\chi(\vec{y},t) + \int_{-\infty}^{t} dt'\chi(\vec{y},t')\vec{\nabla}\cdot\hat{\vec{J}}(\vec{y},t') \tag{60}$$

Substitute this into Eq. (58) to obtain

$$\delta \vec{J}_{vac}^{(1)}(\vec{x},t) = i\langle 0 | \left[ \hat{\vec{J}}(\vec{x},t), \int d\vec{y} \int_{-\infty}^{t} dt' \left( \hat{\vec{J}}(\vec{y},t') \cdot \vec{\nabla}\chi(\vec{y},t') + \chi(\vec{y},t')\vec{\nabla}\cdot\hat{\vec{J}}(\vec{y},t') \right) \right] | 0 \rangle$$
$$+ i\langle 0 | \left[ \hat{\vec{J}}(\vec{x},t), \int \hat{\rho}(\vec{y},t)\chi(\vec{y},t)d\vec{y} \right] | 0 \rangle \quad (61)$$

Rearrange terms to obtain

$$\delta \vec{J}_{vac}^{(1)}(\vec{x},t) = i\langle 0 | \left[ \hat{\vec{J}}(\vec{x},t), \int_{-\infty}^{t} dt' \int d\vec{y} \vec{\nabla} \cdot \left( \hat{\vec{J}}(\vec{y},t')\chi(\vec{y},t') \right) \right] | 0 \rangle$$
$$+ i\langle 0 | \left[ \hat{\vec{J}}(\vec{x},t), \int \hat{\rho}(\vec{y},t)\chi(\vec{y},t)d\vec{y} \right] | 0 \rangle \quad (62)$$

Assume reasonable boundary conditions at $|\vec{y}| \to \infty$ so that

$$\int d\vec{y} \vec{\nabla} \cdot \left( \hat{\vec{J}}(\vec{y},t')\chi(\vec{y},t') \right) = 0 \quad (63)$$

Use this to obtain

$$\delta \vec{J}_{vac}^{(1)}(\vec{x},t) = i\langle 0 | \left[ \hat{\vec{J}}(\vec{x},t), \int \hat{\rho}(\vec{y},t)\chi(\vec{y},t)d\vec{y} \right] | 0 \rangle$$
$$= i\int \langle 0 | \left[ \hat{\vec{J}}(\vec{x},t), \hat{\rho}(\vec{y},t) \right] | 0 \rangle \chi(\vec{y},t)d\vec{y} \quad (64)$$

Use Eq. (56) and the fact that $\hat{H}_o | 0 \rangle = 0$ in the above to obtain

$$\delta \vec{J}_{vac}^{(1)}(\vec{x},t) = i\int \langle 0 | \left[ \hat{\vec{J}}(\vec{x}), \hat{\rho}(\vec{y}) \right] | 0 \rangle \chi(\vec{y},t)d\vec{y} = -i\int ST(\vec{y},\vec{x})\chi(\vec{y},t)d\vec{y} \quad (65)$$

Therefore for $\delta \vec{J}_{vac}^{(1)}(\vec{x},t)$ to be zero, for arbitrary $\chi(\vec{y},t)$, the Schwinger term $\langle 0 | \left[ \hat{\rho}(\vec{y}), \hat{\vec{J}}(\vec{x}) \right] | 0 \rangle$ must be zero. However, we have shown in the previous section that this quantity is not zero. Therefore perturbation theory does not produce a gauge invariant result for the vacuum current. This is why regulation is required. It is needed to remove the non-gauge invariant terms that occur due to the fact that the Schwinger term is nonzero.

**VI. Redefining the vacuum state**

The problem that will be addressed in this section is whether it is possible to reformulate Dirac theory in such a way that the Schwinger term will be zero. According to the discussion in Section II in order for the Schwinger term to be zero there must exist quantum states with less energy than the vacuum. This problem was originally addressed in [8] where it was shown how this result could be achieved by modifying the definition of the vacuum state.

Consider, first, the vacuum state $|0\rangle$ as defined in Section III. The top of the negative energy band has an energy of $-m$ and all states with energy less than $-m$ are occupied by a single electron. This definition will be modified as follows: Define the "modified" vacuum state $|0_c\rangle$ as the quantum state in which each negative energy state in the band between $-m$ and $-E_c$ is occupied and negative energy states with energy less than $-E_c$ are unoccupied where $E_c \to \infty$. The state $|0_c\rangle$ is defined by,

$$|0_c\rangle = \prod_{n;\lambda_n \in \text{band}} \hat{a}_n^\dagger |0, \text{bare}\rangle \tag{66}$$

where the notation $n;\lambda_n \in \text{band}$ means the product is taken over all states in the band of states whose energy is between $-m$ and $-E_c$. This can also be expressed as,

$$\begin{aligned}
&\hat{a}_n |0_c\rangle = 0 \text{ for } \lambda_n = 1 \\
&\hat{a}_n^\dagger |0_c\rangle = 0 \text{ for } -m \geq \lambda_n E_{\vec{p}_n} \geq -E_c \\
&\hat{a}_n |0_c\rangle = 0 \text{ for } -E_c > \lambda_n E_{\vec{p}_n}
\end{aligned} \tag{67}$$

Note the state $|0_c\rangle$ is almost identical to $|0\rangle$ with the exception that for $|0_c\rangle$ the bottom of the negative energy band is handled by a limiting process. The cutoff energy $-E_c$ is assumed to finite and is taken to negative infinity at the end of a calculation. States with less energy then $-E_c$ are unoccupied. If one of these states becomes occupied then the new state will have less free field energy then $|0_c\rangle$. Therefore $|0_c\rangle$ is no longer the lower bound to the free field energy and, as we will show below, if $|0_c\rangle$ is used as the vacuum state then the Schwinger term is zero.

To evaluate the Schwinger term, using $|0_c\rangle$, replace $|0\rangle$ with $|0_c\rangle$ in (42) to obtain,

$$ST_c(\vec{y},\vec{x}) = \langle 0_c|[\hat{\rho}(\vec{y}),\hat{J}(\vec{x})]|0_c\rangle \tag{68}$$

Refer to (44) to yield,

$$ST_c(\vec{y},\vec{x}) = q^2 \sum_{nmrs} \langle 0_c|[\hat{a}_n^\dagger \hat{a}_m, \hat{a}_r^\dagger \hat{a}_s]|0_c\rangle \left(\phi_n^\dagger(\vec{y})\phi_m(\vec{y})\right)\left(\phi_r^\dagger(\vec{x})\vec{\alpha}\phi_s(\vec{x})\right) \tag{69}$$

Use (45) in the above to obtain,

$$ST_c(\vec{y},\vec{x}) = q^2 \sum_{nmrs} \langle 0_c|\begin{pmatrix}\delta_{mr}\hat{a}_n^\dagger \hat{a}_s \\ -\delta_{ns}\hat{a}_r^\dagger \hat{a}_m\end{pmatrix}|0_c\rangle \left(\phi_n^\dagger(\vec{y})\phi_m(\vec{y})\right)\left(\phi_r^\dagger(\vec{x})\vec{\alpha}\phi_s(\vec{x})\right) \tag{70}$$

Use (67) in the above and redefine some of the dummy variables to yield,

$$ST_c(\vec{y},\vec{x}) = q^2 \sum_{s;\lambda_s \in \text{band}} \sum_m \left\{\begin{aligned}&\left(\phi_s^\dagger(\vec{y})\phi_m(\vec{y})\right)\left(\phi_m^\dagger(\vec{x})\vec{\alpha}\phi_s(\vec{x})\right) \\ &-\left(\phi_s^\dagger(\vec{x})\vec{\alpha}\phi_m(\vec{x})\right)\left(\phi_m^\dagger(\vec{y})\phi_s(\vec{y})\right)\end{aligned}\right\} \tag{71}$$

The notation $s;\lambda_s \in \text{band}$ means the index 's' is summed over the states whose energy is in the band from $-m$ to $-E_c$. Note that the summation over 'm' is over all states.

Take the summation over 'm' and use (38) in the above to obtain,

$$\mathrm{ST}_c(\vec{y},\vec{x}) = q^2 \sum_{s;\lambda_s \in \mathrm{band}} \left\{ \begin{array}{l} \left(\phi_s^\dagger(\vec{y})\vec{\alpha}\phi_s(\vec{x})\right)\delta^{(3)}(\vec{x}-\vec{y}) \\ -\left(\phi_s^\dagger(\vec{x})\vec{\alpha}\phi_s(\vec{y})\right)\delta^{(3)}(\vec{x}-\vec{y}) \end{array} \right\} \tag{72}$$

Next use the relationship,

$$f(\vec{y})\delta^{(3)}(\vec{x}-\vec{y}) = f(\vec{x}) \tag{73}$$

to obtain,

$$\mathrm{ST}_c(\vec{y},\vec{x}) = q^2 \sum_{s;\lambda_s \in \mathrm{band}} \delta^{(3)}(\vec{x}-\vec{y}) \left\{ \begin{array}{l} \left(\phi_s^\dagger(\vec{x})\vec{\alpha}\phi_s(\vec{x})\right) \\ -\left(\phi_s^\dagger(\vec{x})\vec{\alpha}\phi_s(\vec{x})\right) \end{array} \right\} = 0 \tag{74}$$

Therefore the Schwinger term is zero if vacuum state is defined in such away that there is no lower bound to the free field energy. If this is done the theory will be mathematically consistent. In addition perturbation theory will produce a gauge invariant result without the need for regularization.

## **VII. Summary and Conclusion**

In Section I we introduced four elements that are normally considered part of Dirac field theory. Using the first three elements (the Schrödinger equation, the continuity equation, and the principle of gauge invariance) it was shown that the Schwinger term is zero. However, when the fourth element (that there is a lower bound to the free field energy) is used the Schwinger term is nonzero. Therefore Dirac theory is not mathematically consistent.

In Section III the vacuum state vector $|0\rangle$ is defined according to the standard definition [5]. It is shown the Schwinger term is non-zero in this case. This result was expected because $|0\rangle$ is defined in such a way as to be the lower bound to the free field energy. The problem with this result is that the theory will not be gauge invariant and the continuity equation will not hold.

This inconsistency shows up when the vacuum current is calculated using perturbation theory. In Section V it was shown that if the Schwinger term is nonzero then perturbation theory yields a nongauge invariant result for the vacuum current. This is why regularization is necessary in Dirac field theory. In Section VI it was shown how the definition of the vacuum state can be modified so that the Schwinger term is zero. The modified vacuum state $|0_c\rangle$ differs from the conventional vacuum state $|0\rangle$ in that the bottom of the negative energy band is defined using a limiting procedure. This allows for the theoretical existence of quantum states with less energy than that of the vacuum state. This is a necessary condition for the Schwinger term to be zero.

If $|0_c\rangle$ is used as the vacuum state then Dirac theory will be mathematically consistent and perturbation theory will produce a gauge invariant vacuum current without the need for regularization.